\pdfoutput=1
\documentclass[preprint,preprintnumbers,amsmath,amssymb,floatfix,endfloats*]{revtex4}
\usepackage{graphicx}% Include figure files
\usepackage{dcolumn}% Align table columns on decimal point
\usepackage{bm}% bold math
\usepackage{amsfonts}

\DeclareMathAlphabet{\mathsfsl}{OT1}{cmr}{bx}{it}
\begin{document}
%----------------------------------------------------------------------%
% Title
%----------------------------------------------------------------------%
%
\title{Relaxation dynamics in amorphous alloys under asymmetric cyclic shear deformation}
\author{Pritam Kumar Jana$^{1}$ and Nikolai V. Priezjev$^{2}$}
\affiliation{$^{1}$Department of Chemistry, Birla Institute of
Technology and Science, Pilani, Pilani Campus, Rajasthan 333031,
India}
\affiliation{$^{2}$Department of Mechanical and Materials
Engineering, Wright State University, Dayton, OH 45435}
\date{\today}
\begin{abstract}

The influence of cyclic loading and glass stability on structural
relaxation and yielding transition in amorphous alloys was
investigated using molecular dynamics simulations. We considered a
binary mixture cooled deep into the glass phase and subjected to
cyclic shear deformation where strain varies periodically but
remains positive. We found that rapidly cooled glasses under
asymmetric cyclic shear gradually evolve towards states with lower
potential energy and finite stress at zero strain. At the strain
amplitude just below a critical value, the rescaled distributions of
nonaffine displacements converge to a power-law decay with an
exponent of about -2 upon increasing number of cycles. By contrast,
more stable glasses yield at lower strain amplitudes, and the
yielding transition can be delayed for hundreds of cycles when the
strain amplitude is near a critical value. These results can be
useful for the design of novel thermo-mechanical processing methods
to improve mechanical and physical properties of metallic glasses.

\vskip 0.5in


\end{abstract}

\maketitle

\section{Introduction}

The development and optimization of bulk metallic glasses is
important for various structural and biomedical applications
including wear-resistant gears~\cite{Hofmann17} and biocompatible
implants~\cite{Loye21}. It is well known that due to their amorphous
structure, metallic glasses possess advantageous properties such as
high strength, large elastic strain limit, and high corrosion
resistance~\cite{Egami13}.  Under applied strain, the elementary
plastic deformation in such amorphous alloys involve swift
rearrangement of a small group of atoms or shear
transformations~\cite{Spaepen77, Argon79}. Upon further loading,
well annealed glasses typically yield via the formation of narrow
shear bands where strain is localized~\cite{Shi19}. It was shown
that the ductile behavior can be restored by relocating glasses to
higher energy states~\cite{Greer16}. Such rejuvenation can be
achieved via high-pressure torsion, cold rolling, wire drawing,
irradiation, elastostatic loading, and other thermo-mechanical
processing methods~\cite{Greer16}. A particularly elegant method
involves cryogenic thermal cycling that can rejuvenate metallic
glasses due to plastic deformation under internal stresses arising
during spatially heterogeneous thermal expansion~\cite{Ketov15,
Guo19,Priez18tcyc,Priez19T2000,Priez19T5000,Mirdamadi19,Priez19one,
Jittisa20,Zhang20,Tjahjono21,Guan21,Varnik22,Amigo22}. Moreover, the
process of rejuvenation during elastostatic compression can be
accelerated when stress is temporarily applied along alternating
directions~\cite{PriezELAST19,PriezELAST21}. In addition, it was
also recently demonstrated experimentally and by means of molecular
dynamics (MD) simulations that higher energy states can be realized
via rapid freezing of a glass former under applied
stress~\cite{Mota21,PriezMETAL21}. However, in spite of the progress
made, the design of novel strategies for thermo-mechanical
processing of metallic glasses is required to access a broader range
of energy states and improved mechanical properties.

\vskip 0.05in

In the last decade, the dynamic response of amorphous materials to
cyclic loading was extensively investigated using atomistic
simulations~\cite{Priezjev13,Reichhardt13,Sastry13,Hoy13,
IdoNature15,Priezjev16,Kawasaki16,Priezjev16a,Sastry17,Priezjev17,
OHern17, Priezjev18, Priezjev18a, NVP18strload, Sastry19band,
PriezSHALT19, ShuoLi19, Priez20ba, Jana20, KawBer20, NVP20altY,
Priez20delay, BhaSastry21, Krishnan21, Priez21var, Priez21heal,
PriezCMS21, Regev22, PriezJNCS22, Sastry22a, Peng22, Sastry22}.
Interestingly, it was demonstrated that in a broad range of energy
states, athermal glasses under oscillatory quasistatic shear
deformation exhibit a sharply defined yielding transition; and, in
general, cyclic shear provides a better characterization of the
yielding transition than uniform loading~\cite{Sastry17}. Moreover,
the yielding transition in well annealed binary glasses can be
delayed for thousands of cycles when temperature is well below
$T_g$~\cite{Priez20delay}, whereas at about half $T_g$, the critical
strain amplitude remains the same for stable glasses with initially
different energy levels~\cite{PriezJNCS22}. It was further shown
that under small-amplitude cyclic loading, rapidly quenched glasses
undergo structural relaxation, termed as `mechanical annealing', and
gradually evolve towards low-energy steady states~\cite{Sastry13,
Priezjev18,Sastry19band,PriezSHALT19,Jana20}. In some cases,
mechanically induced annealing can be accelerated when the shear
orientation is periodically alternated in two or three spatial
dimensions~\cite{PriezSHALT19,Krishnan21} or if the strain amplitude
of cyclic loading is occasionally increased above a critical
value~\cite{Priez21var}.  More recently, it was found that the
yielding behavior in binary glasses under asymmetric cyclic shear
deformation is markedly different from the case of symmetric
loading, and, in particular, better annealed glasses under
asymmetric shear yield at a smaller strain amplitude and exhibit an
enhanced plastic activity at strain amplitudes below a critical
value~\cite{Sastry22a}. Despite extensive modeling efforts, however,
the influence of the deformation protocol and preparation history on
the yielding transition and mechanical annealing remain not fully
understood.

\vskip 0.05in

In this paper, we present results of molecular dynamics simulations
of binary glasses subjected to asymmetric oscillatory deformation
where shear strain varies periodically but remains positive. It will
be shown that rapidly cooled glasses under cyclic loading at strain
amplitudes below a critical value evolve towards progressively lower
energy states with a finite stress at zero strain. This relaxation
process proceeds via irreversible rearrangements of atoms whose
nonaffine displacements become power-law distributed.  We also find
that stable glasses undergo a yielding transition at smaller strain
amplitudes and the onset of yielding is delayed for hundreds of
shear cycles near the critical strain amplitude.

\vskip 0.05in

The rest of this paper is organized as follows. We describe the
details of molecular dynamics simulations and the asymmetric
deformation protocol in the next section. The analysis of the
potential energy, nonaffine displacements, and stress-strain
response is presented in section\,\ref{sec:Results}. The results are
briefly summarized in the last section.

\section{Molecular dynamics simulations}
\label{sec:MD_Model}

In our study, the amorphous alloy was represented by means of the
binary (80:20) Lennard-Jones (LJ) mixture model originally proposed
by Kob and Andersen (KA) over twenty years ago~\cite{KobAnd95}. In
the KA model, the interaction between atoms of types
$\alpha,\beta=A,B$ is specified via the LJ potential, as follows:
\begin{equation}
V_{\alpha\beta}(r)=4\,\varepsilon_{\alpha\beta}\,\Big[\Big(\frac{\sigma_{\alpha\beta}}{r}\Big)^{12}\!-
\Big(\frac{\sigma_{\alpha\beta}}{r}\Big)^{6}\,\Big],
\label{Eq:LJ_KA}
\end{equation}
where the parameters are set to $\varepsilon_{AA}=1.0$,
$\varepsilon_{AB}=1.5$, $\varepsilon_{BB}=0.5$, $\sigma_{AA}=1.0$,
$\sigma_{AB}=0.8$, $\sigma_{BB}=0.88$, and
$m_{A}=m_{B}$~\cite{KobAnd95}. A similar parameters were employed by
Weber and Stillinger to study the amorphous metal-metalloid alloy
$\text{Ni}_{80}\text{P}_{20}$~\cite{Weber85}. In the present study,
the results are reported in terms of the reduced units of length,
mass, energy, and time, as follows: $\sigma=\sigma_{AA}$, $m=m_{A}$,
$\varepsilon=\varepsilon_{AA}$, and
$\tau=\sigma\sqrt{m/\varepsilon}$. The total number of atoms is
$N=60\,000$ and the cutoff radius is
$r_{c,\,\alpha\beta}=2.5\,\sigma_{\alpha\beta}$. The MD simulations
were performed using the LAMMPS code with the integration time step
$\triangle t_{MD}=0.005\,\tau$~\cite{Allen87,Lammps}.

\vskip 0.05in

% equilibration and temperature protocol

\begin{figure}[t]
\includegraphics[width=12.0cm,angle=0]{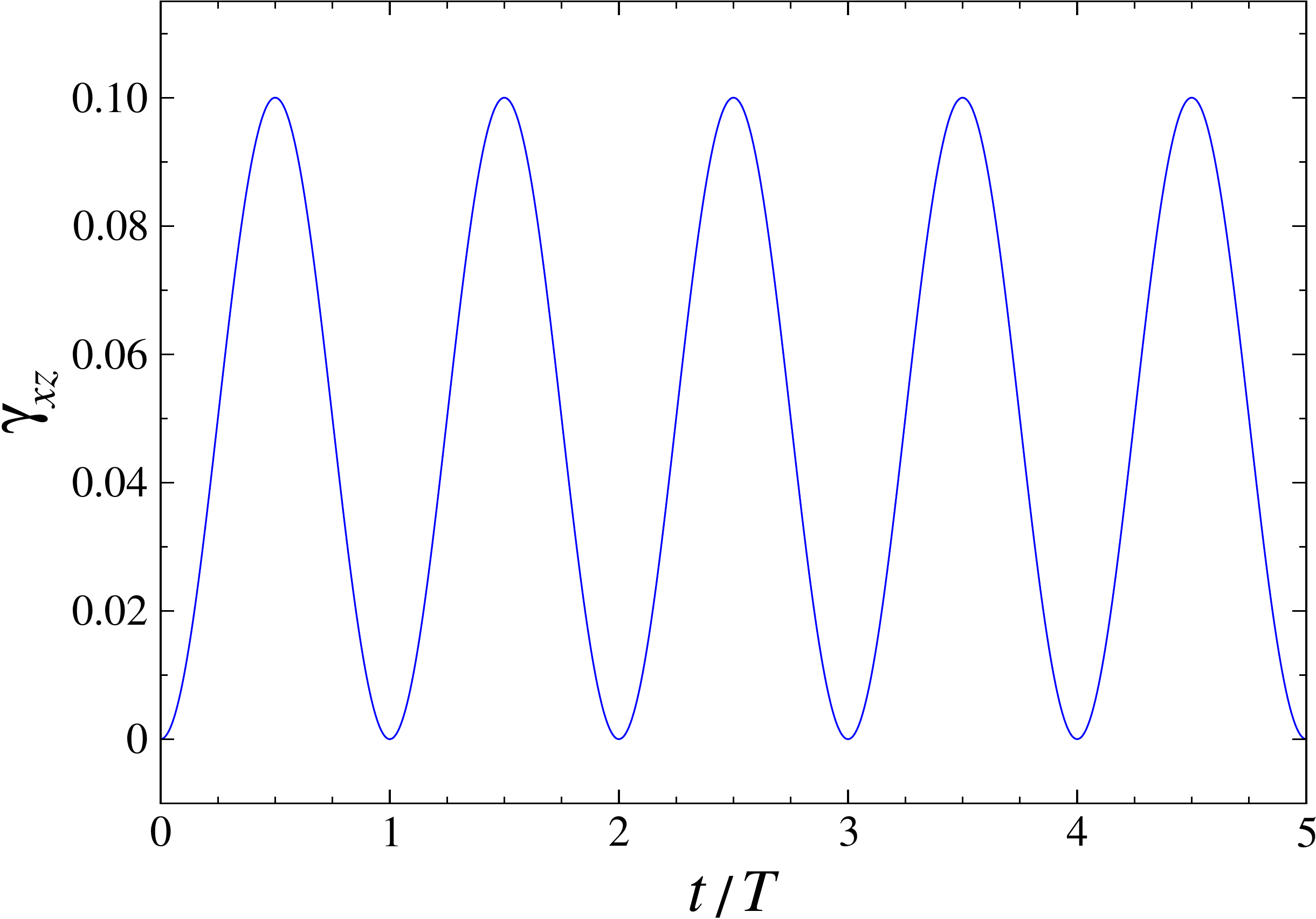}
\caption{The applied shear strain, $\gamma_{xz}$, as a function of
time, $t/T$, specified by Eq.\,(\ref{Eq:shear}) for the strain
amplitude $\gamma_0=0.10$.  The oscillation period is
$T=5000\,\tau$. }
\label{fig:abs_strain}
\end{figure}

The sample preparation procedure involved the following steps.
First, the binary mixture was placed in a cubic box of size
$L=36.84\,\sigma$ and equilibrated at the temperature
$T_{LJ}=1.0\,\varepsilon/k_B$ and density
$\rho=\rho_A+\rho_B=1.2\,\sigma^{-3}$. Here, $k_B$ is the Boltzmann
constant. This temperature is well above the critical temperature
$T_g=0.435\,\varepsilon/k_B$ of the KA model at the density
$\rho=1.2\,\sigma^{-3}$~\cite{KobAnd95}. In all simulations, the
temperature was controlled via the the Nos\'{e}-Hoover
thermostat~\cite{Allen87,Lammps} and periodic boundary conditions
were applied along the $\hat{x}$, $\hat{y}$, and $\hat{z}$
directions. Second, the system was rapidly cooled from the
temperature $T_{LJ}=1.0\,\varepsilon/k_B$ to $0.01\,\varepsilon/k_B$
with the rate of $10^{-2}\varepsilon/k_{B}\tau$ at constant density
$\rho=1.2\,\sigma^{-3}$. After the thermal quench, the binary glass
was subjected to time periodic shear deformation along the $xz$
plane, as follows:
\begin{equation}
\gamma_{xz}(t)=\gamma_0\,[1+\text{sin}(2\pi t/T - \pi/2)]/2,
\label{Eq:shear}
\end{equation}
where $\gamma_0$ is the strain amplitude, $T=5000\,\tau$ is the
oscillation period, and the strain $\gamma_{xz}$ varies from zero to
$\gamma_0$ and then back to zero during one period. The example of
the imposed shear strain for the strain amplitude $\gamma_0=0.10$
during five oscillation periods is shown in
Fig.\,\ref{fig:abs_strain}. The production run involved 1000 shear
cycles for each value of $\gamma_0$ in the range from $0.04$ to
$0.13$ at $T_{LJ}=0.01\,\varepsilon/k_B$ and
$\rho=1.2\,\sigma^{-3}$. The simulations were performed for one
sample due to computational limitations, and the potential energy,
stress components, and atomic configurations were periodically saved
for post-processing analysis.

\section{Results}
\label{sec:Results}

% intro and brief review

It has long been realized that mechanical and physical properties of
metallic glasses depend sensitively on the rate of cooling from the
liquid state~\cite{Greer16}. In particular, more slowly cooled
glasses typically settle at lower energy states and upon large
enough deformation might exhibit brittle yielding via strain
localization along narrow bands. By contrast, rapid cooling across
the glass transition temperature generally results in higher energy
states and ductile behavior~\cite{Greer16}. Moreover, metallic
glasses can be further relaxed to lower energy states via
oscillatory deformation when the strain amplitude is below a
critical value~\cite{Sastry13,Priezjev18,PriezSHALT19}. In what
follows, we consider rapidly and slowly cooled binary glasses under
asymmetric cyclic deformation, where the imposed shear strain is
varied periodically but remains positive, and compare the simulation
results with the case of symmetric deformation with respect to zero
strain.

\vskip 0.05in

% potential energy during asymmetric cyclic loading; poorly annealed glass

\begin{figure}[t]
\includegraphics[width=12.0cm,angle=0]{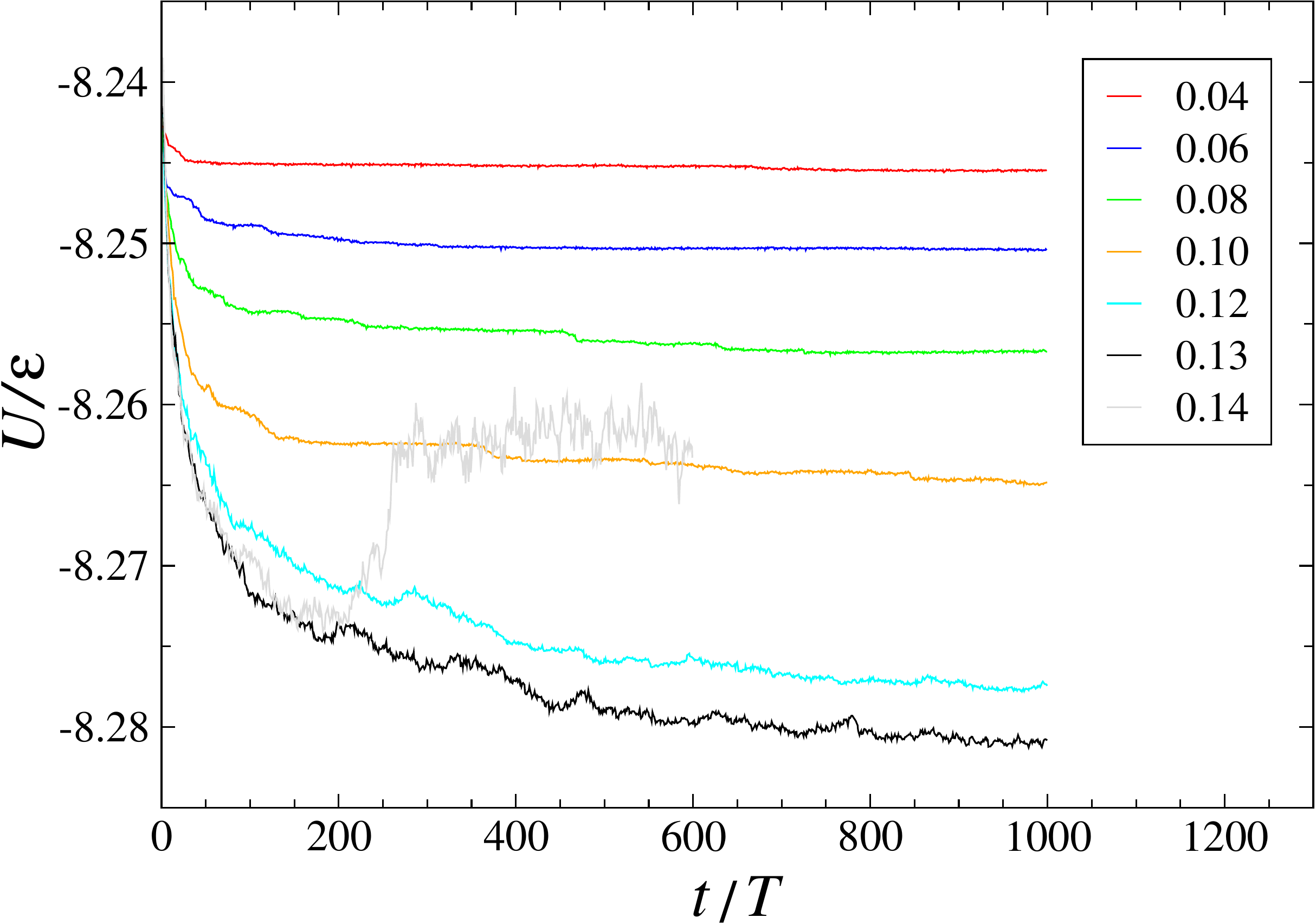}
\caption{(Color online) The potential energy minima as a function of
the number of shear cycles for the indicated values of strain
amplitude $\gamma_0$ in Eq.\,(\ref{Eq:shear}). The period of
oscillation is $T=5000\,\tau$. The glass was initially prepared by
cooling across the glass transition temperature with the rate of
$10^{-2}\varepsilon/k_{B}\tau$ to $T_{LJ}=0.01\,\varepsilon/k_B$ at
$\rho=1.2\,\sigma^{-3}$. }
\label{fig:poten_cyc_anneal}
\end{figure}

In our analysis, a rapidly quenched binary glass was repeatedly
loaded for $1000$ cycles using the asymmetric shear deformation
protocol given by Eq.\,(\ref{Eq:shear}). The potential energy minima
during each cycle are shown in Fig.\,\ref{fig:poten_cyc_anneal} for
the indicated values of the strain amplitude. As is evident, the
glass is relocated to lower energy states upon increasing strain
amplitude. Note that, as loading continues, the potential energy
levels off to a plateau for $\gamma_0 \leqslant 0.10$, whereas
progressively lower energy states are attained for $0.12 \leqslant
\gamma_0 \leqslant 0.13$. These conclusions are consistent with the
results of the previous study of rapidly quenched glasses under
symmetric cyclic shear deformation~\cite{PriezSHALT19,Priez21var}.
For example, the asymmetric loading at $\gamma_0=0.12$ results in
the same energy level, $U \approx -8.277\,\varepsilon$, as the
symmetric cyclic shear deformation at the strain amplitude of $0.06$
for the same operating conditions~\cite{PriezSHALT19}. As shown in
Fig.\,\ref{fig:poten_cyc_anneal}, mechanically induced annealing at
$\gamma_0=0.14$ proceeds for about 200 cycles, followed by the
yielding transition via the formation of a shear band with a higher
energy. Similar yielding behavior was reported for symmetric cyclic
shear deformation of rapidly quenched binary glasses at the strain
amplitude of $0.07$~\cite{NVP20altY}.

\vskip 0.05in

% potential energy during the last 3 cycles

\begin{figure}[t]
\includegraphics[width=12.0cm,angle=0]{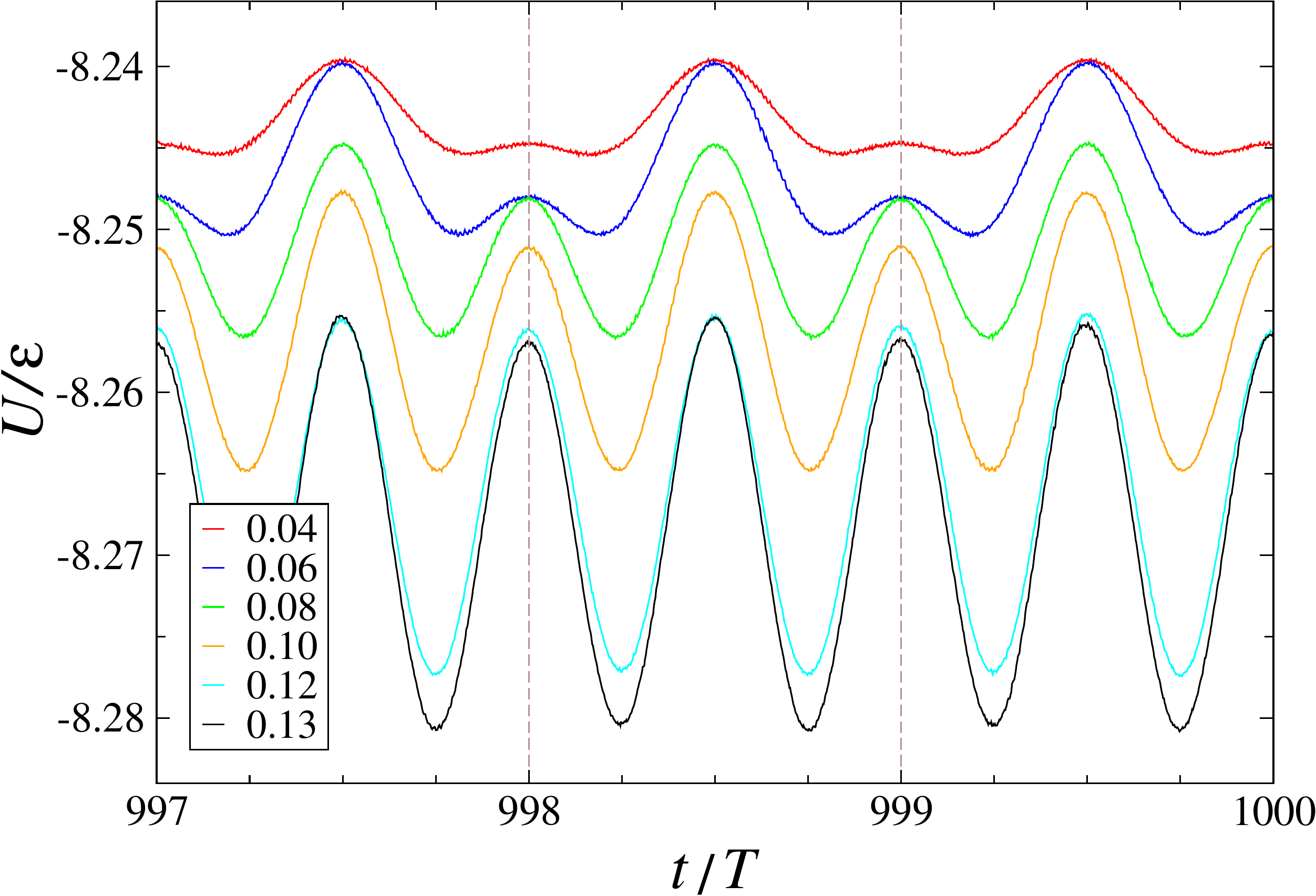}
\caption{(Color online) The variation of the potential energy during
the last three loading cycles for the indicated strain amplitudes
$\gamma_0$. The oscillation period is $T=5000\,\tau$. The same data
as in Fig.\,\ref{fig:poten_cyc_anneal}. The vertical dashed lines
are plotted for reference. }
\label{fig:poten_last3cyc}
\end{figure}

The main difference between symmetric and asymmetric cyclic shear
deformation of rapidly quenched glasses is that in the latter case
the potential energy minima as well as zero stress are attained at
nonzero strain. For example, the variation of the potential energy
during the last three cycles is shown in
Fig.\,\ref{fig:poten_last3cyc} for the strain amplitudes $\gamma_0
\leqslant 0.13$. It can be seen that for $0.08 \leqslant \gamma_0
\leqslant 0.13$, the potential energy reaches minima at about
$0.25\,T$ and $0.75\,T$ when strain is $\gamma_0/2$, while for
smaller strain amplitudes, $\gamma_0 = 0.04$ and $0.06$, the energy
minima become less pronounced and occur at lower strain. These
results imply that for sufficiently large strain amplitudes, rapidly
quenched glasses under asymmetric cyclic shear deformation evolve
towards symmetric deformation with respect to $\gamma_0/2$.

\vskip 0.05in

% definition of D2min

The spatial and temporal distribution of local plastic events in
amorphous alloys can be quantified via the analysis of the so-called
nonaffine displacements of atoms~\cite{Falk98}. By definition, the
nonaffine measure for the $i$-th atom is calculated using the
transformation matrix $\mathbf{J}_i$, which linearly transforms a
group of neighboring atoms and, at the same time, minimizes the
following expression:
\begin{equation}
D^2(t, \Delta t)=\frac{1}{N_i}\sum_{j=1}^{N_i}\Big\{
\mathbf{r}_{j}(t+\Delta t)-\mathbf{r}_{i}(t+\Delta t)-\mathbf{J}_i
\big[ \mathbf{r}_{j}(t) - \mathbf{r}_{i}(t)  \big] \Big\}^2,
\label{Eq:D2min}
\end{equation}
where the sum is taken over $N_i$ neighbors that are located within
$1.5\,\sigma$ from the position of the $i$-th atom,
$\mathbf{r}_{i}(t)$, and $\Delta t$ is the time interval between two
configurations of atoms. The local plastic rearrangement is
typically associated with the values of the nonaffine measure
$D^2(t, \Delta t)$ greater than the cage size, which for the KA
model at $\rho=1.2\,\sigma^{-3}$ is about
$0.1\,\sigma^2$~\cite{KobAnd95}. It was recently shown, however,
that during cyclic shear deformation below the yielding transition,
a fraction of atoms undergo reversible nonaffine displacements with
amplitudes that are approximately power-law
distributed~\cite{Priezjev16,Priezjev16a}.

\vskip 0.05in

% distributions of D^2

\begin{figure}[t]
\centering
\includegraphics[width=0.49\linewidth]{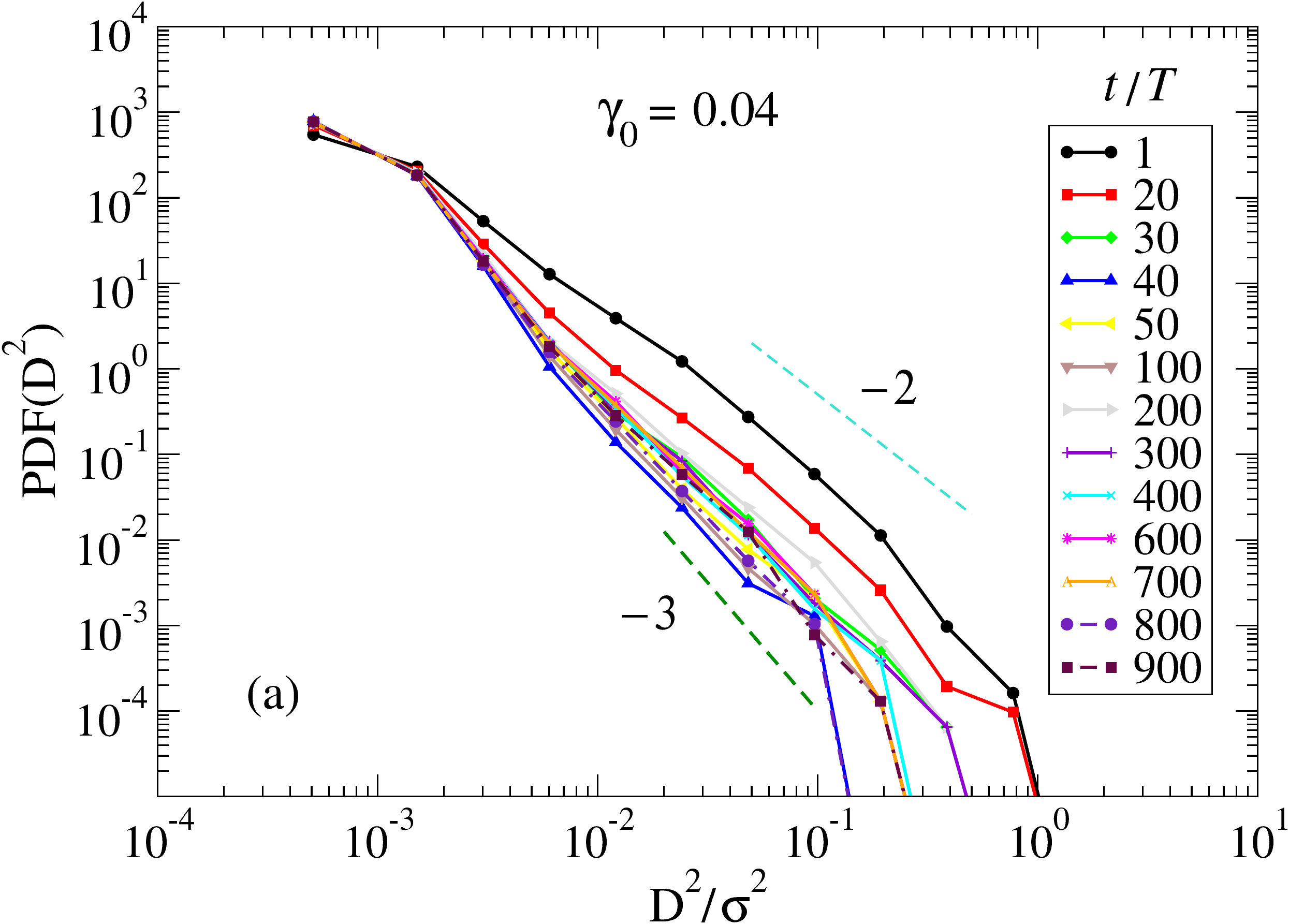}
\includegraphics[width=0.49\linewidth]{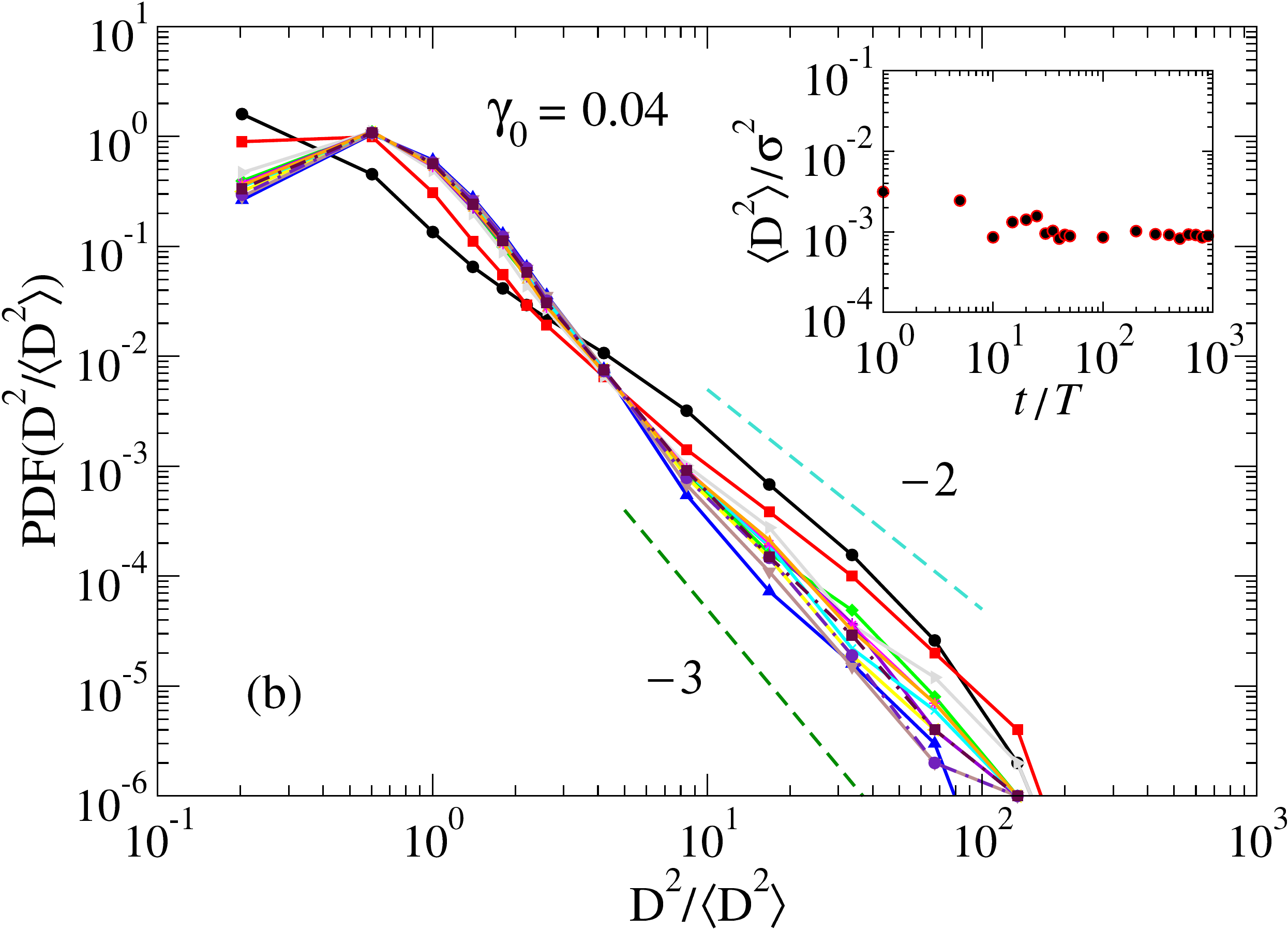}

\vspace{0.8 cm}

\includegraphics[width=0.49\linewidth]{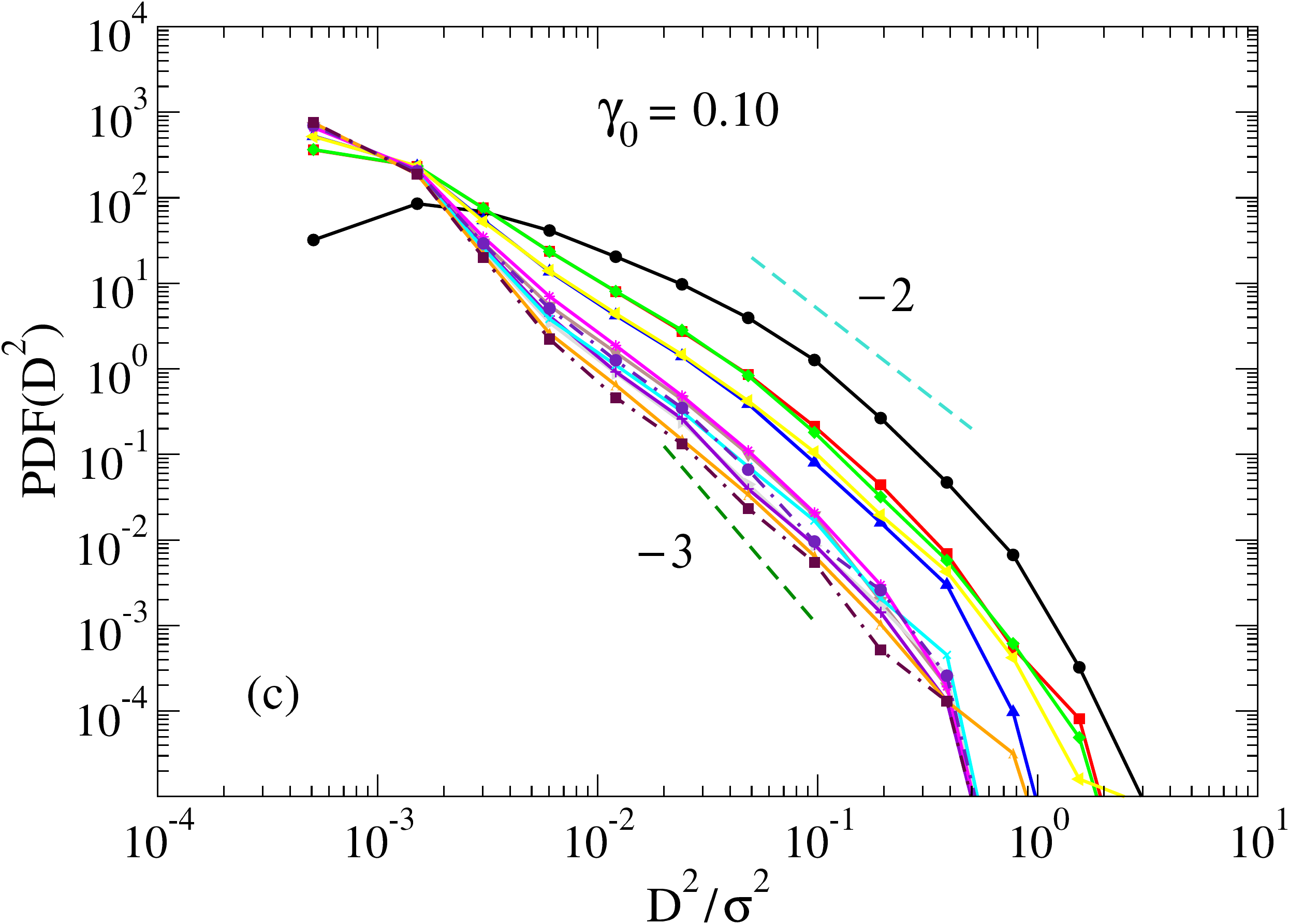}
\includegraphics[width=0.49\linewidth]{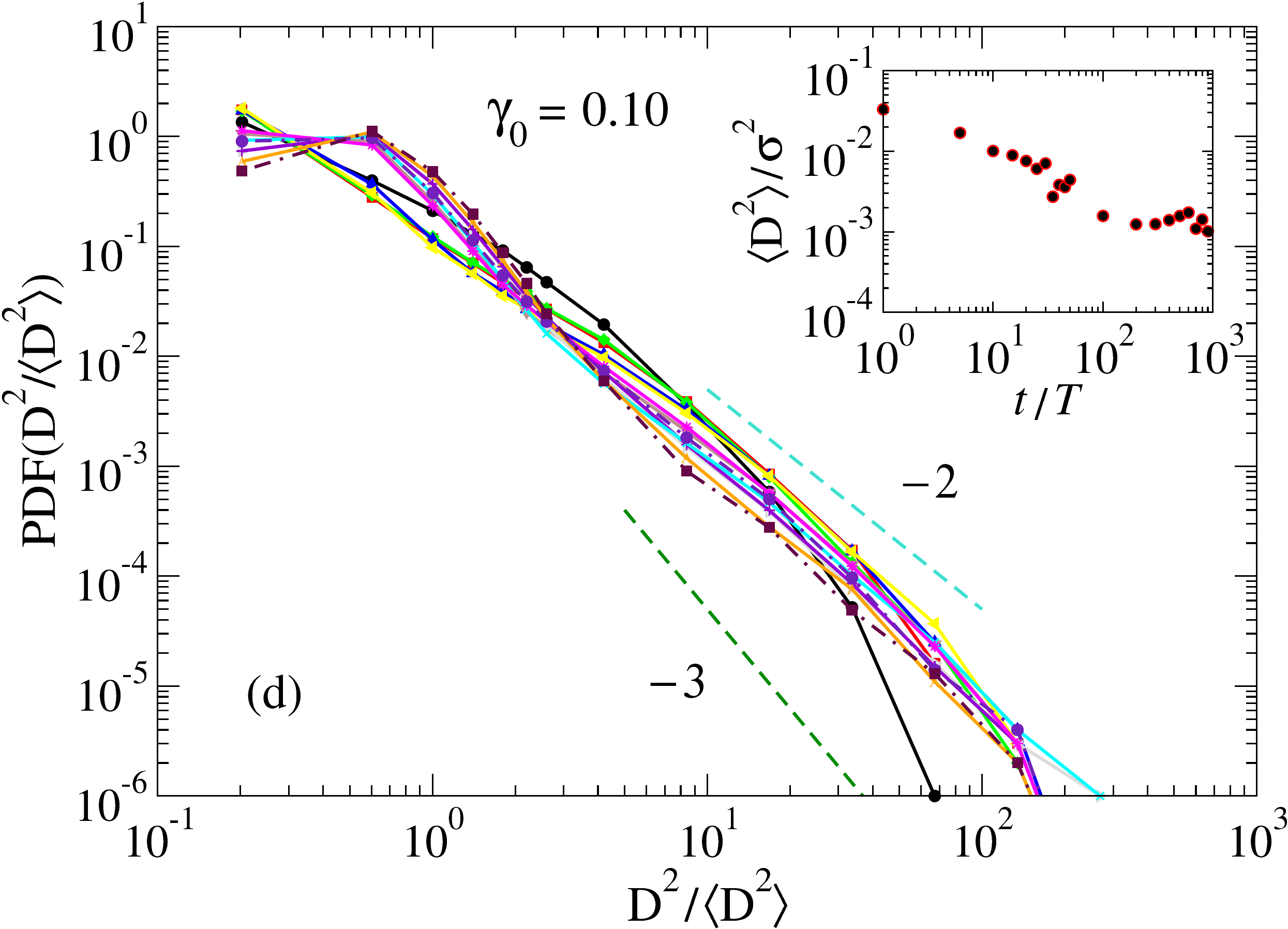}

\vspace{0.8 cm}

\includegraphics[width=0.49\linewidth]{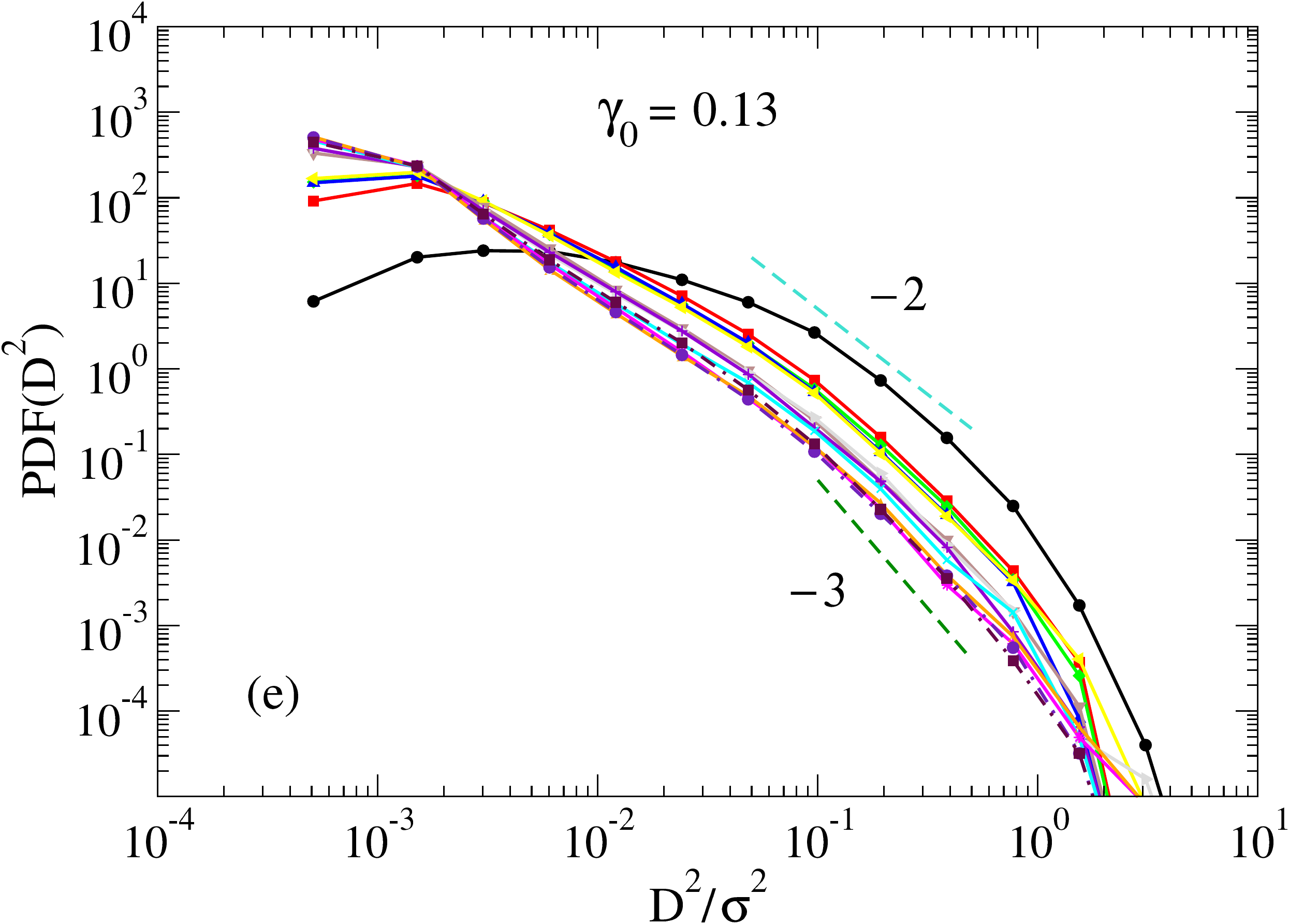}
\includegraphics[width=0.49\linewidth]{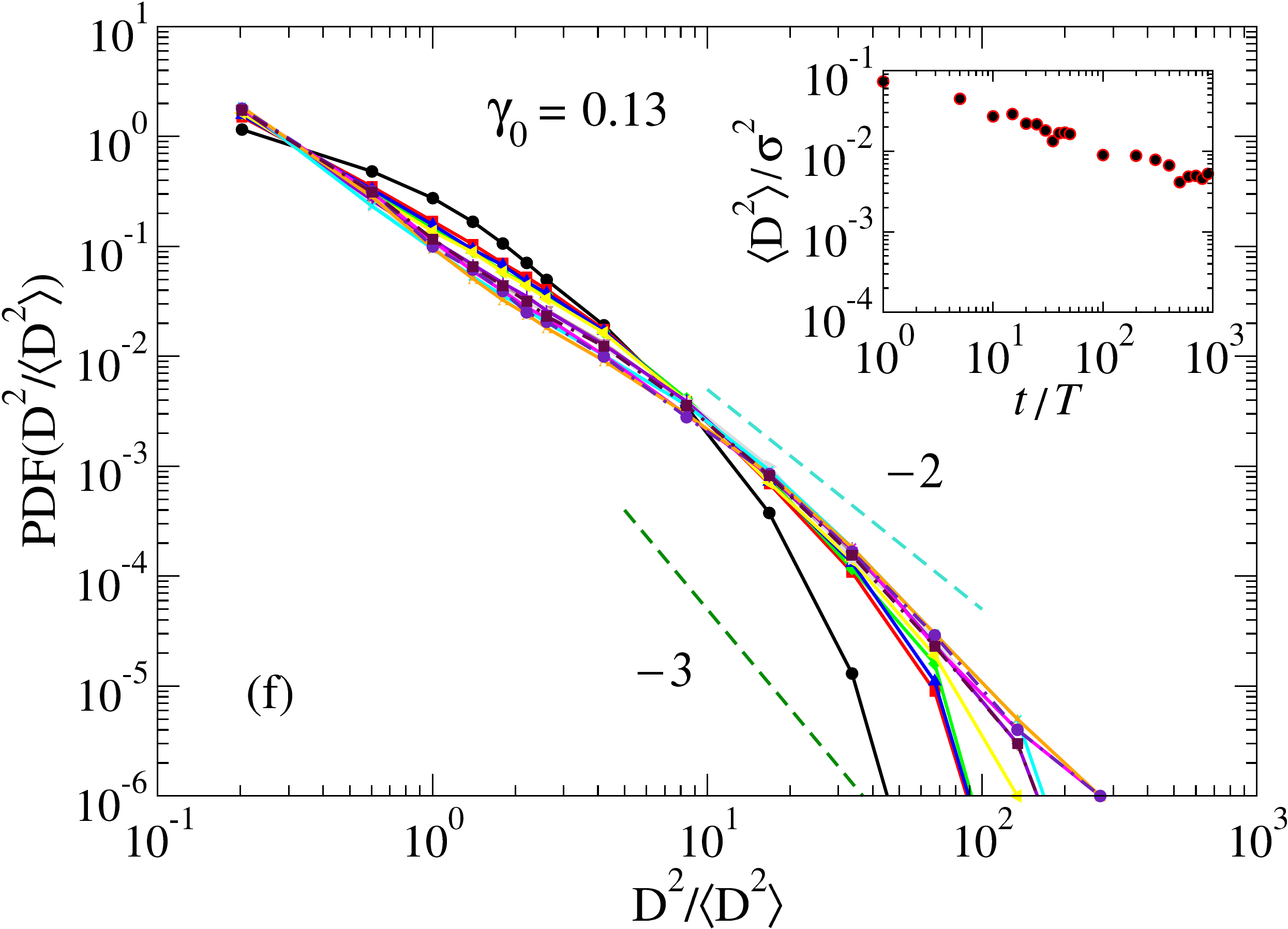}
\caption{The probability distributions of the nonaffine quantity,
$D^2(t, \Delta t = T)$, (left panels) and the corresponding rescaled
distributions (right panels) for (a, b) $\gamma_0=0.04$, (c, d)
$0.10$, and (e, f) $0.13$. The cycle numbers, $t/T$, are listed in
the legend. The insets show the average values of $D^2$ as a
function of $t/T$. The dashed lines with the slopes of -2 and -3 are
plotted for reference. }
\label{fig:probD2}
\end{figure}

The probability distribution functions of the nonaffine measure are
presented in Figs.\,\ref{fig:probD2}a, \ref{fig:probD2}c, and
\ref{fig:probD2}e for the strain amplitudes $\gamma_0=0.04$, $0.10$,
and $0.13$, respectively. The nonaffine quantity $D^2(t, \Delta t)$
in Eq.\,(\ref{Eq:D2min}) was computed for atomic configurations at
zero strain at the beginning and end of a cycle, i.e., $\Delta t =
T$, for the tabulated values of the cycle number $t/T$ shown in
Fig.\,\ref{fig:probD2}a. Note also that the data in
Fig.\,\ref{fig:probD2} are presented on the log-log scale using
logarithmic bin widths to avoid noise at large values of $D^2$. It
can be clearly seen that distributions for $t/T=1$ are relatively
wide and extend up to $D^2\approx 1.0\,\sigma^2$, indicating
enhanced plastic activity at the beginning of the loading process.
With increasing cycle number, the distributions of $D^2$ tend to
become more narrow but remain approximately power-law distributed.
Notice, however, that in some cases, more narrow distributions
correspond to smaller number of loading cycles, e.g., $t/T=40$ in
Fig.\,\ref{fig:probD2}a, which can be attributed to fluctuations in
the average of $D^2$ as a function of the cycle number (see insets
in the right panels of Fig.\,\ref{fig:probD2}).

\vskip 0.05in

% distributions of D^2 continued

Further, the corresponding rescaled distributions of $D^2/\langle
D^2 \rangle$ are presented in Figs.\,\ref{fig:probD2}b,
\ref{fig:probD2}d, and \ref{fig:probD2}f for the same strain
amplitudes $\gamma_0=0.04$, $0.10$, and $0.13$. It can be seen in
Fig.\,\ref{fig:probD2}f that for the largest strain amplitude
$\gamma_0=0.13$, the distributions of $D^2/\langle D^2 \rangle$
converge to a single curve with the power-law exponent of about $-2$
upon increasing number of cycles.  This behavior might be related to
a continued structural relaxation at $\gamma_0=0.13$ during $1000$
cycles, which is reflected in the fast decay of the potential energy
reported in Fig.\,\ref{fig:poten_cyc_anneal} and relatively large
values of $\langle D^2 \rangle$ shown in the inset of
Fig.\,\ref{fig:probD2}f. The accelerated relaxation at
$\gamma_0=0.13$ involves a large number of plastic events and thus
results in broad distributions of $D^2$ during $1000$ cycles. By
contrast, for the lower strain amplitudes, $\gamma_0=0.04$ and
$0.10$, the deformation becomes nearly reversible after a number of
transient cycles (see Fig.\,\ref{fig:poten_cyc_anneal}). As a
result, the average of $D^2$ is much smaller than the typical cage
size, and the distributions of $D^2/\langle D^2 \rangle$ become
weakly dependent on the cycle number due to relatively large scatter
in $\langle D^2 \rangle$, as shown in Figs.\,\ref{fig:probD2}b and
\ref{fig:probD2}d.

\vskip 0.05in

% shear stress

\begin{figure}[t]
\includegraphics[width=12.0cm,angle=0]{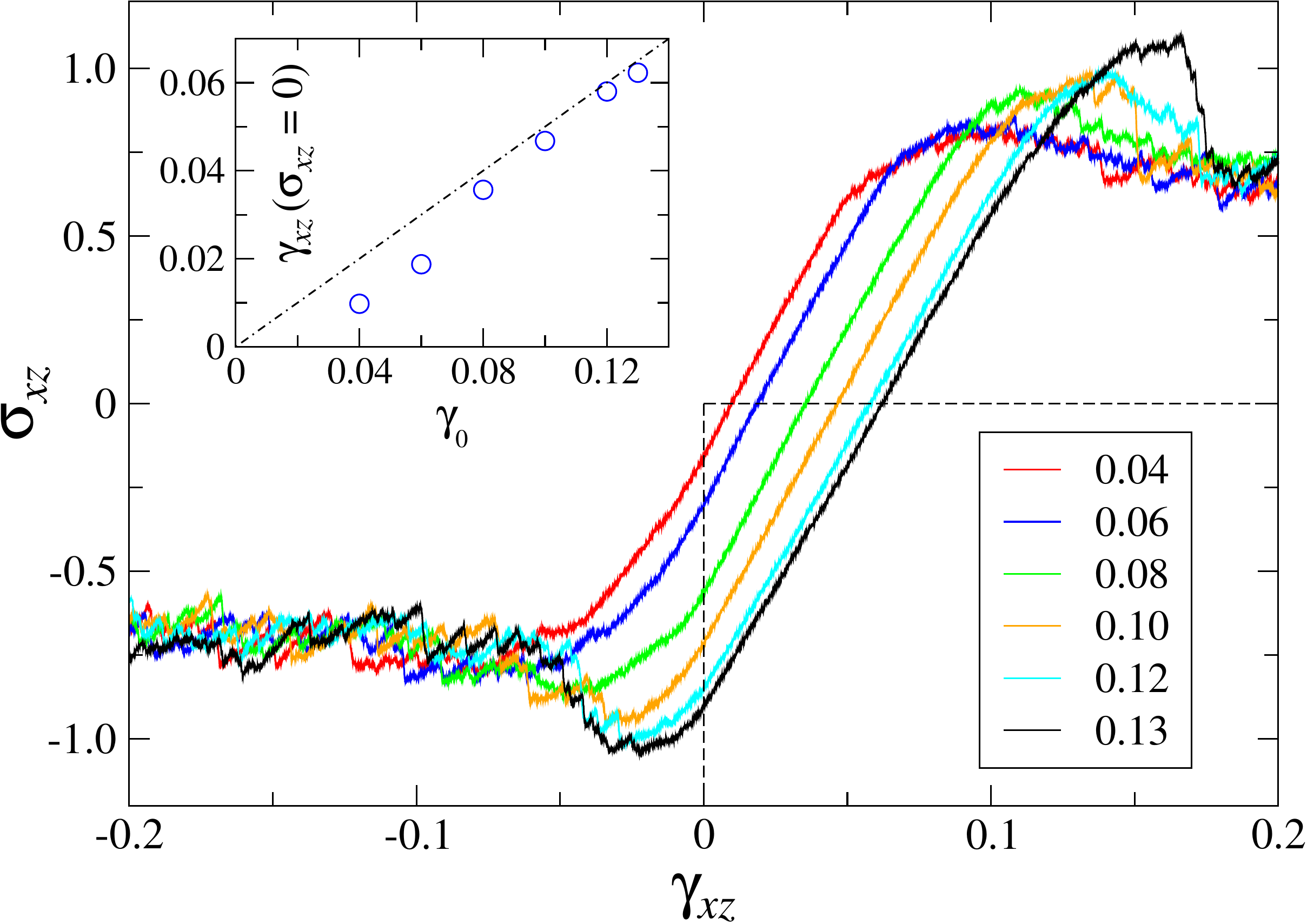}
\caption{(Color online) The shear stress, $\sigma_{xz}$ (in units of
$\varepsilon\sigma^{-3}$), as a function of strain, $\gamma_{xz}$,
during steady loading with the rate of $10^{-5}\tau^{-1}$. The
values of the strain amplitude, $\gamma_0$, used for mechanical
annealing are listed in the legend. The dashed lines indicate
$\gamma_{xz}=0$ and $\sigma_{xz}=0$. The inset shows shear strain as
zero stress versus the strain amplitude. The dash-dotted line
$y=0.5\,x$ is plotted for reference.}
\label{fig:steady_stress_strain}
\end{figure}

After periodic deformation for $1000$ shear cycles, the binary glass
was sheared at a constant strain rate of $10^{-5}\tau^{-1}$ from
zero strain to 0.2 and to -0.2. The results for the shear stress as
a function of strain are displayed in
Fig.\,\ref{fig:steady_stress_strain} for the indicated values of the
strain amplitude used for cyclic loading. It can be seen that with
increasing $\gamma_0$ (lower energy glasses), the magnitude of
stress at zero strain as well as the peak value of the stress
overshoot increase. The maximum value of the yielding peak is
$\sigma_{y}\approx1.1\,\varepsilon\sigma^{-3}$ for the better
annealed sample at $\gamma_0=0.13$. In turn, the inset in
Fig.\,\ref{fig:steady_stress_strain} shows the shear strain at zero
stress as a function of the strain amplitude of cyclic loading. Note
that except for the cases $\gamma_0=0.04$ and $0.06$, the shear
strain $\gamma_{xz}(\sigma_{xz}=0)$ is about half the strain
amplitude, indicating that during cyclic loading, the oscillation of
shear stress becomes nearly symmetric with respect to $\gamma_0/2$.
For all values of $\gamma_0$, the relative strain
$\gamma_{xz}(\sigma_{xz}=0)/\gamma_0$ correlates well with the
locations of minima in the potential energy reported in
Fig.\,\ref{fig:poten_last3cyc}. These results are in agreement with
a recent MD study on binary glasses under asymmetric cyclic
deformation using athermal quasistatic shear
protocol~\cite{Sastry22a}.

\vskip 0.05in

% number of atoms with large nonaffine displacements

\begin{figure}[t]
\includegraphics[width=12.0cm,angle=0]{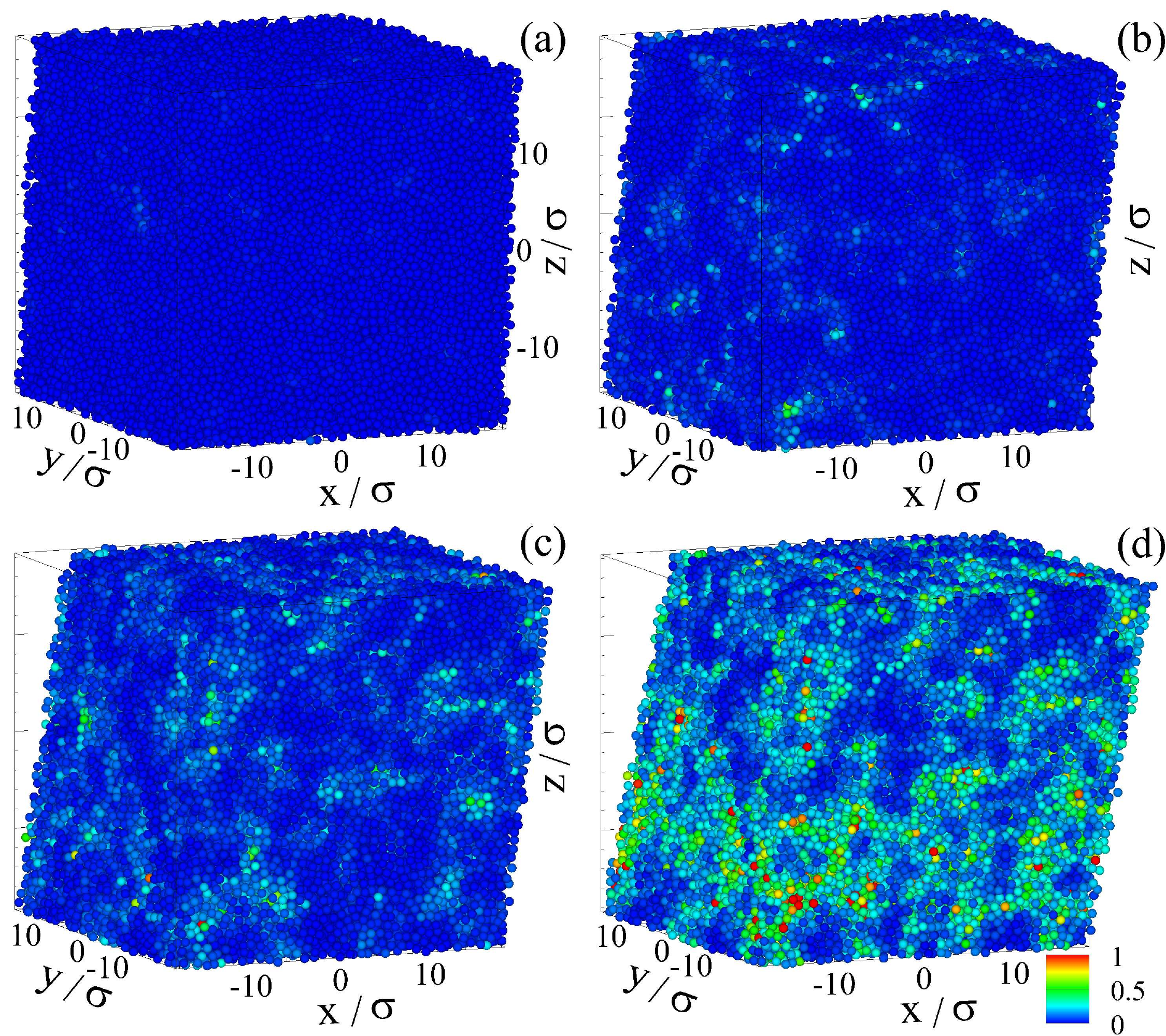}
\caption{(Color online) The atomic configurations of the binary
glass strained along the \textit{xz} plane with the strain rate of
$10^{-5}\tau^{-1}$. The shear strain is (a) 0.05, (b) 0.10, (c)
0.15, and (d) 0.20. The colorcode indicates values of $D^2$ computed
with respect to the atomic configuration at zero strain. The glass
was initially cooled across $T_g$ with the rate of
$10^{-2}\varepsilon/k_{B}\tau$ and periodically deformed for 1000
shear cycles with $\gamma_0=0.04$ in Eq.\,(\ref{Eq:shear}).}
\label{fig:snapshot_amp020_1000_p_D2}
\end{figure}

\begin{figure}[t]
\includegraphics[width=12.0cm,angle=0]{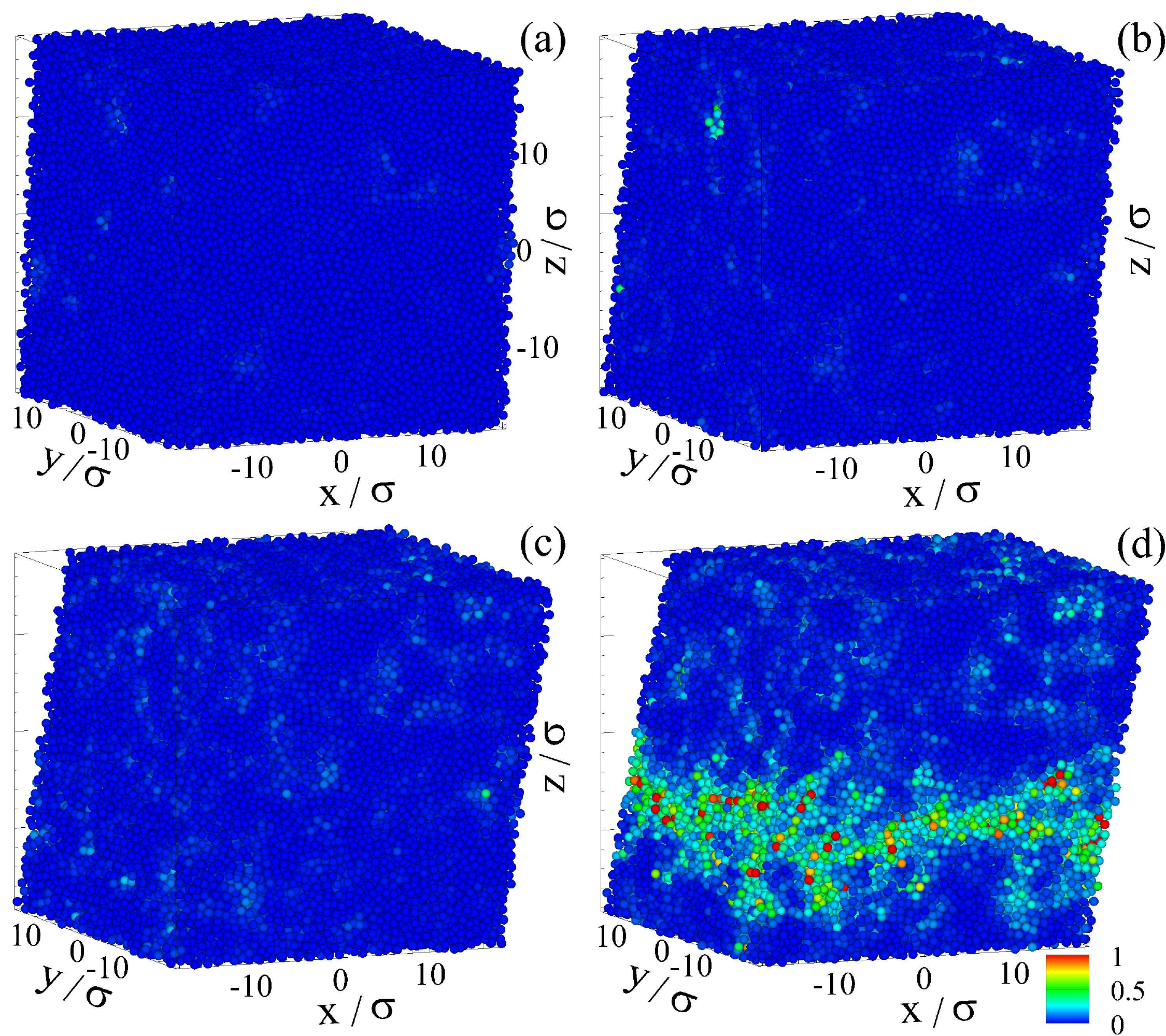}
\caption{(Color online) The snapshots of strained glass along the
\textit{xz} plane with the rate of $10^{-5}\tau^{-1}$. The strain is
(a) 0.05, (b) 0.10, (c) 0.15, and (d) 0.20. The nonaffine measure
$D^2$ is indicated by the color in the legend. The sample was
initially cooled with the rate of $10^{-2}\varepsilon/k_{B}\tau$ and
loaded for 1000 shear cycles with the strain amplitude
$\gamma_0=0.13$.}
\label{fig:snapshot_amp065_1000_p_D2}
\end{figure}

The shear stress response to the applied strain shown in
Fig.\,\ref{fig:steady_stress_strain} indicates that a glass
undergoes a transition from ductile to brittle failure upon
increasing strain amplitude of cyclic loading. To illustrate
microscopic details of plastic deformation, we presented a series of
snapshots at shear strain ranging from 0.05 to 0.2 in
Fig.\,\ref{fig:snapshot_amp020_1000_p_D2} for the glass initially
annealed at the strain amplitude $\gamma_0=0.04$ and in
Fig.\,\ref{fig:snapshot_amp065_1000_p_D2} for $\gamma_0=0.13$. In
each panel, the nonaffine measure was computed with respect to an
atomic configuration at zero strain. It can be clearly observed in
Fig.\,\ref{fig:snapshot_amp020_1000_p_D2} that the yielding
transition in the poorly annealed sample is associated with
homogeneous distribution of atoms with large nonaffine
displacements. In contrast, the yielding transition in the more
stable glass annealed at $\gamma_0=0.13$ proceeds via the formation
of a shear band, as shown in
Fig.\,\ref{fig:snapshot_amp065_1000_p_D2}. Interestingly, nonaffine
displacements in the better annealed glass remain relatively small
when strain is varied in the range
$0\leqslant\gamma_{xz}\leqslant0.15$, despite a large variation in
shear stress of about twice the yielding peak (see the black curve
in Fig.\,\ref{fig:steady_stress_strain}).

\vskip 0.05in

% well annealed, stable glasses

\begin{figure}[t]
\includegraphics[width=12.0cm,angle=0]{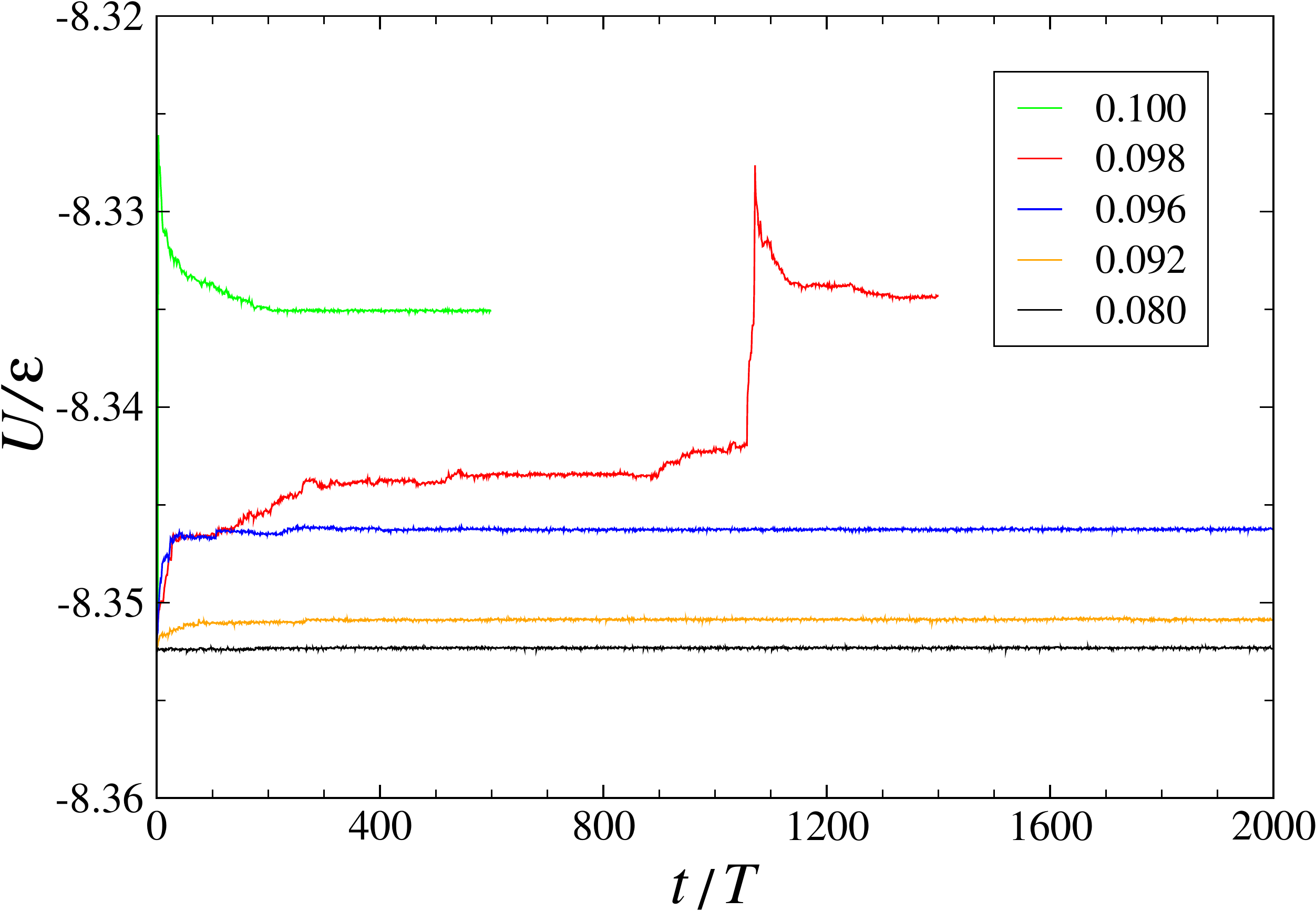}
\caption{(Color online) The potential energy minima in periodically
deformed stable glasses for the tabulated values of the strain
amplitude $\gamma_0$ in Eq.\,(\ref{Eq:shear}). The oscillation
period is $T=5000\,\tau$. }
\label{fig:stable_glass}
\end{figure}

Lastly, we include a comment about the yielding transition in stable
glasses under asymmetric cycle shear deformation. In this case, the
binary mixture was first slowly cooled from
$T_{LJ}=1.0\,\varepsilon/k_B$ to $0.3\,\varepsilon/k_B$ with the
rate of $10^{-5}\varepsilon/k_{B}\tau$ at $\rho=1.2\,\sigma^{-3}$,
then periodically deformed for 5000 cycles, and finally cooled to
$T_{LJ}=0.01\,\varepsilon/k_B$ at $\rho=1.2\,\sigma^{-3}$. In the
previous study~\cite{PriezCMS21}, it was shown that this preparation
procedure leads to a relatively stable glass with the potential
energy $U\approx -8.352\,\varepsilon$ in the undeformed state.
Figure\,\ref{fig:stable_glass} shows the energy minima during each
cycle in the stable glass loaded at the selected strain amplitudes.
It can be seen that the yielding transition, which is reflected in
an abrupt energy change, occurs after loading is applied at
$\gamma_0=0.100$, while at $\gamma_0 \leqslant 0.096$, after about
300 cycles, the potential energy remains nearly constant for $2000$
cycles, indicating reversible deformation. For $\gamma_0 = 0.092$
and $0.096$, the energy increase during the first 300 cycles is
associated with plastic deformation, similar to the results reported
for athermal glasses under asymmetric shear~\cite{Sastry22a}.
Furthermore, loading at $\gamma_0=0.098$ results in a stepwise
increase of the potential energy during 1000 cycles and the yielding
transition after 1057 cycles (see the red curve in
Fig.\,\ref{fig:stable_glass}). This behavior is consistent with the
delayed yielding transition in stable glasses sinusoidally deformed
at a finite temperature~\cite{Priez20delay}. Altogether, these
results demonstrate that the yielding transition in more stable
glasses under the deformation protocol given by
Eq.\,(\ref{Eq:shear}) occurs at lower strain amplitudes and the
transition can be significantly delayed when the strain amplitude is
in the vicinity of the critical value.

\section{Conclusions}

In summary, we investigated the effects of asymmetric cyclic shear
deformation and preparation history of amorphous alloys on
mechanical annealing and yielding transition using molecular
dynamics simulations. We considered a binary mixture rapidly cooled
deep into the glass phase and applied periodic deformation where
shear strain varied sinusoidally but remained positive. Similar to
the symmetric case, we demonstrated that cyclic deformation induced
structural relaxation and resulted in a gradual decay of the
potential energy upon continued loading. It was further shown that
rescaled distributions of nonaffine displacements during one cycle
followed a power-law decay when the strain amplitude is below a
critical value. In agreement with the recent results for athermal
glasses under asymmetric cyclic shear~\cite{Sastry22a}, we found
that the yielding transition in more stable glasses occurs at lower
strain amplitudes. In the presence of thermal fluctuations, the
yielding transition can be delayed for hundreds of shear cycles in a
relatively large system when periodically deformed at strain
amplitudes near a critical value.

\section*{Acknowledgments}

Financial support from the National Science Foundation (CNS-1531923)
is gratefully acknowledged. Molecular dynamics simulations were
performed at Wright State University's Computing Facility and the
Ohio Supercomputer Center using the LAMMPS code~\cite{Lammps}.

% \section*{Conflict of Interest}

%%%%%%%%%%%%%%% FIGURES %%%%%%%%%%%%%%%%%%%%%%%

% definition of the modulated strain amplitude
%

% potential energy minima during each cycle
%

% potential energy during the last 3 cycles
%

% distributions of D^2
%

% stress strain after 1000 cycles
%

% steady strain snapshots ka_60_T001_T5000_shear_xz_p_amp020_1000_p_D2
%

% steady strain snapshots ka_60_T001_T5000_shear_xz_p_amp065_1000_p_D2
%

% yielding in stable glass
%

\bibliographystyle{prsty}

\end{document}